\documentclass[11pt,a4paper]{article}

\usepackage{a4wide}
\usepackage{amssymb}
\usepackage{xcolor}
\usepackage{hyperref}
\usepackage{amsmath}

\usepackage{amsfonts}
\usepackage{amsmath}
\usepackage{amssymb}
\usepackage{bm}
\usepackage{dcolumn}
\usepackage{epsfig}
\usepackage{graphicx}
\usepackage{graphics}
\usepackage[latin1]{inputenc}
\usepackage{latexsym}
\usepackage{rotating}
\usepackage{hyperref}
\usepackage{dsfont}

\usepackage{color}
\usepackage[active]{srcltx}
\usepackage{amsmath,amsfonts,amssymb,amsthm,amstext,amscd,eucal,srcltx}
\usepackage{epsfig,graphicx,bm}
\usepackage{epstopdf, epsf}
\usepackage{dcolumn}
\usepackage{hyperref}

\newcommand{\bea}{\begin{eqnarray}}
\newcommand{\eea}{\end{eqnarray}}

\usepackage{amssymb} \usepackage{amsmath} \usepackage{graphicx}
\usepackage{epsfig,latexsym}

\def\bef{\begin{figure}}
\def\eef{\end{figure}}

\newcommand{\be}[1]{\begin{equation}\label{#1}}
\newcommand{\beq}{\begin{equation}}
\newcommand{\ee}{\end{equation}}
\newcommand{\beqn}[1]{\begin{eqnarray}\label{#1}}
\newcommand{\eeqn}{\end{eqnarray}}
\newcommand{\bd}{\begin{displaymath}}
\newcommand{\ed}{\end{displaymath}}

\def\lsim{\raise0.3ex\hbox{$\;<$\kern-0.75em\raise-1.1ex
e\hbox{$\sim\;$}}}
\def\gsim{\raise0.3ex\hbox{$\;>$\kern-0.75em\raise-1.1ex
\hbox{$\sim\;$}}}
\def\simlt{\mathrel{\lower2.5pt\vbox{\lineskip=0pt\baselineskip=0pt
           \hbox{$<$}\hbox{$\sim$}}}}
\def\simgt{\mathrel{\lower2.5pt\vbox{\lineskip=0pt\baselineskip=0pt
           \hbox{$>$}\hbox{$\sim$}}}}
\def\unity{{\hbox{1\kern-.8mm l}}}

\def\beq{\begin{equation}}
\def\eeq{\end{equation}}

\def\lsim{\mathrel{\mathop  {\hbox{\lower0.5ex\hbox{$\sim$}
\kern-0.8em\lower-0.7ex\hbox{$<$}}}}}
\def\gsim{\mathrel{\mathop  {\hbox{\lower0.5ex\hbox{$\sim$}
\kern-0.8em\lower-0.7ex\hbox{$>$}}}}}

\begin{document}
\begin{center}
{\LARGE Conformal Bootstrap in dS/CFT and Topological Quantum Gravity}\\
\vspace{1.5em}

\vspace{1.5em}

Andrea Addazi\footnote{e-mail address: {\tt andrea.addazi@qq.com}}
and Antonino Marcian\`o\footnote{e-mail address: {\tt marciano@fudan.edu.cn}}\\
\vspace{0.5em}
Department of Physics \& Center for Field Theory and Particle Physics,\\
Fudan University, 200433 Shanghai, China

\end{center}

%
%

\vspace*{3mm}
\date{\today}

\begin{abstract}
\noindent
We show that the correspondence among $AdS_{3}/CFT_{2}$, the 1D Schwarzian Model, Sachdev-Ye-Kitaev model and 2+1D Topological Quantum Gravity can be extended to the case of  $dS_{3}/CFT_{2}$. The $R$-matrix, related to the gravitational scattering amplitude near the horizon of $dS_{3}$ black hole, corresponds (on the side of the holographic projection) to a crossing kernel in the Schwarzian Model. The $R$-matrix is related to the 6j-symbol of SU$(1,1)$. We also find that in the Euclidean $dS_{3}$ a new Kac-Moody symmetry of instantons emerges out. We dub these new solutions {\it Kac-Moodions}. A one-to-one correspondence of Kac-Moodion levels and SU$(2)$ spin representations is established. Every instanton then corresponds to spin representations deployed in Topological Quantum Gravity. The instantons are directly connected to the Black Hole entropy, as punctures on its horizon. This strongly supports the recent proposal, in arXiv:1707.00347, that a Kac-Moody symmetry of gravitational instantons is related to the black hole information processing. We also comment on a further correspondence that can be established between the Schwarzian Model and non-commutative spacetimes in 2+1D, passing through the equivalence with Topological Quantum Gravity with cosmological constant, in the limit when the latter vanishes. 
\end{abstract}


\section{Introduction}

Recently, there has been much interest in the literature in trying to unify different aspects of string theory and quantum gravity models with the holographic principle and the $AdS/CFT$ correspondence. In particular, a new holographic connection involving the $AdS_{3}/CFT_{2}$ correspondence, the 1D Sachdev-Ye-Kitaev (SYK), the Schwarzian Model and 2+1D Topological Quantum Gravity (TQG) was discovered and analyzed in Refs.~\cite{Turiaci:2017zwd,Mertens:2017mtv}. 

Following a similar pathway, Strominger proposed in Ref.~\cite{Strominger:2001pn} a $dS/CFT$ correspondence, as an extension of the holographic principle contained in Maldacena's seminal proposal \cite{Maldacena:1997re}. This urgently rises a further question: {\it is it possible to extend the series of correspondences mentioned above to the context of dS/CFT?}

In this paper, we will show that also within context of $dS_{3}/CFT_{2}$ a correspondence with TQG can be established. Furthermore, we show that a correspondence that can be also established between the Schwarzian Model and non-commutative spacetimes in 2+1D. The latter deploys the limit of vanishing cosmological constant in 2+1D Topological Quantum Gravity and the recover of non-commutative space-times.

\section{Linking $dS_{3}/CFT_{2}$ with the Schwarzian Model}

We spell in this section a series of relations that deeply connects the Schwarzian Model in quantum mechanics to the $dS_{3}/CFT_{2}$ correspondence. 

\subsection{Asymptotic symmetries of $dS_{3}$}

Let us start considering the $dS_{3}$ metric, which reads 
\be{jjj}
ds^{2}=-\left(-\frac{r^{2}}{l^{2}}+1\right)dt^{2}+\left(-\frac{r^{2}}{l^{2}}+1\right)^{-1}dr^{2}+r^{2}d\theta^{2}\,.
\ee
This is exactly the same metric of $AdS_3$, but with the $l^{2}\rightarrow -l^{2}$ replacement --- $l$ is the Hubble radius of the metric. As renown, the $dS_{3}$ metric has six Killing vectors, namely the doublet of getenerators 
\be{kkl}
J_{0}^{\pm}=-\frac{1}{2}(l\partial_{t}\pm \imath \partial_{\theta}),
\ee
and the quadruplets of generators 
\be{Jpm}
J_{\sigma=\pm}=\frac{1}{2}e^{\sigma(t\mp \imath l\theta)/l}\sqrt{r^{2}-l^{2}}\Big(\sigma \partial_{r}-\frac{rl}{r^{2}-l^{2}}\partial_{t}\mp \imath \frac{1}{r}\partial_{\theta}\Big)\, . 
\ee
The Killing vectors generate the $\mathfrak{sl}(2,\mathbb{C})$ algebra  
\be{jkl}
[J_{1}^{+},J_{-1}^{+}]=2J_{0}^{+},\,\,\, [J_{0}^{+},J_{\pm}^{+}]=\mp J_{\pm 1}^{+}
\ee
where $(J_{n}^{+})^{*}=J_{n}^{-}$.



The asymptotic symmetry of $dS_{3}$ is characterized by one Killing vector, which reads
\be{vector}
\zeta=U\partial_{z}+\frac{1}{2}e^{2t}U''\partial_{\bar{z}}+\frac{1}{2}U'\partial_{t}\, , 
\ee
where $U=U(z)$ is a holomorphic function and $'$ stands for the derivative with respect to $z=x_1+\imath x_2$, the space-coordinates being denoted with $x_1$ and $x_2$. The metric transforms under a diffeomorphism according to the Lie derivative
\be{jk}
\delta_{\zeta}g_{mn}=-\mathcal{L}_{\zeta}g_{mn}\,.
\ee
In terms of the $U$-parametrization the Eq.~\eqref{vector} recasts 
\be{jjj}
\delta_{U}g_{zz}=-\frac{l^{2}}{2}U'''\, , 
\ee
and 
\be{kk}
\delta_{U}g_{z\bar{z}}=\delta_{U}g_{zt}=\delta_{U}g_{tt}=0\, . 
\ee
The metric is invariant under the Killing transformation $\zeta$, if and only if $U'''$ is vanishing. A generic solution of $U(z)$ reads $U=\alpha+\beta z+\gamma z^{2}$. These transformations correspond to the conformal group of the complex plane, while the isometry group corresponds to the SL$(2,\mathbb{C})$ subgroup of the asymptotic symmetry group.\\

The iso-group of $AdS_{3}$ is SO$(2,2)$, which in the Euclidean rotation $t\rightarrow \imath t$ allows for an isomorphic identification with SL$(2,\mathbb{R})\times$SL$(2,\mathbb{R})$. In stead, the iso-group of $dS_{3}$ is SO$(3,1)$, the double covering of which is SL$(2,\mathbb{C})$. Since the group is already SO$(3,1)$, which contains a double copy of SL$(2,\mathbb{R})$, the Euclidean continuation is not required. In stead, the Wick rotation to the Euclidean space entails the SO$(4)$ group, which corresponds to SU$(2)\times$SU$(2)$. In this latter case, no reparametrization freedom, such as M\"obius transformations, are present, but a rotation freedom.

\subsection{The Schwarzian Model}

Since the symmetry of $dS_{3}$ is SL$(2,\mathbb{R})\times$ SL$(2,\mathbb{R})$, an holographic connection to the SL$(2,\mathbb{R})$ invariant 1D Schwarzian model can be envisaged. Let us introduce the Schwarzian Model, with the action cast
\be{SYK}
S[f]=-C\int_{0}^{\beta}d\tau\{F,\tau\}\,,
\ee
\be{hhj}
\{F,\tau\}=\{f,\tau \}+\frac{2\pi^{2}}{\beta^{2}}f'^{2}\,,\qquad F=\tan\left(\frac{\pi f(\tau)}{\beta}\right)\, , 
\ee
where $C$ is the coupling constant of the (zero-temperature) theory. The function $f$ runs over the (thermal) circle $S_{1}$, while the ${\rm Diff}(S_{1})$ acts on $f(\tau)$ as  $f(\tau+\beta)=f(\tau)+\beta$. This induces the Schwarzian structure
\be{hk}
\{f,\tau\}=\frac{f'''}{f'}-\frac{3}{2}\left(\frac{f''}{f'}\right)^{2}\,,
\ee
which is dubbed Schwarzian derivative. The action in Eq.~\eqref{SYK} is invariant under SL$(2,\mathbb{R})$ M\"obius transformations
\be{jl}
F\rightarrow \frac{aF+b}{cF+d}\, . 
\ee
The Hamiltonian of the model is simply equal to the Casimir of the SL$(2,\mathbb{R})$ group, {\it i.e.} $H=l_{a}l_{a}/2$ where $l_{a}$ stand for the $\mathfrak{sl}(2,\mathbb{R})$ generators. The partition function of the theory is 
\be{hjk}
Z(\beta)=\int_{\mathcal{M}\equiv {\rm Diff}(S_{1})/{\rm SL}(2,\mathbb{R})} \mathcal{D}f e^{-S[f]}\, , 
\ee
where the integration is performed modding out the overall SL$(2,\mathbb{R})$ symmetry. 



The 1D Schwarzian Lagrangian can be related to the complexified 
extended model by means of
\be{jjjkl}
L=\frac{1}{2}\dot{\phi}^{2}+\pi_{f}\dot{f}+\pi_{\bar{f}}\dot{\bar{f}}+\pi_{f}\pi_{\bar{f}}e^{\phi}\,.
\ee
Compared to $AdS$ action, the last term is flipped in sign. In the complexified phase space of theory related to Eq.~\eqref{jjjkl}, the flip of the sign can be achieved by means of the shift $\phi \rightarrow \phi +\imath \pi $.
Starting from Eq.(\ref{jjjkl}), one can obtain the 1D Schwarzian model with the following operations: 
First integrating out $\bar{f}$ and fixing $\pi_{\bar{f}}=\pm 1$, then integrating out $\pi_{f}$. 
The lagrangian obtained is $L=\frac{1}{2}(f''/f')^{2}$, which can be easily rewritten in explicitly 
$SL(2,R)$ invariant form if adding the zero-equivalent global derivative term $-(f''/f)'/2$. 
 
Looking at different limits of Eq.~(\ref{jjjkl}), one may obtain the three models:
{\it i)} 1D Schwarzian quantum mechanics; {\it ii)} 1D Liouville theory; {\it iii)} motion of a particle in the Euclidean $AdS$ or $dS$ backgrounds (depending on the procedure deployed to integrate out modes).\\ 

The 1D Schwarzian model can be obtained by modding out the complex variable $\bar{f}$ and $\pi_{f}$ while fixing the momenta  $\pi_{\bar{f}}=\pm 1$ in Eq.(\ref{jjjkl}). On the other hand, the 1D Liouville model can be obtained from the first form 1D Schwarzian Model from integrating out $f$ and fixing the momenta $\pi_{f}=\pm \mu=\pm {\rm const}$, namely 
\be{kla}
L=\frac{1}{2}\dot{\phi}^{2}\pm \mu e^{\phi}\, . 
\ee

The motion of a particle on the Euclidean $AdS_{3}(dS_{3})$ space can be achieved by integrating out $\pi_{\bar{f}}$ as it follows:
\be{LLLK}
L=\frac{1}{2}\dot{\phi}^{2}\pm \dot{f}\dot{\bar{f}}e^{-\phi}\,. 
\ee
The sign difference arrises from the different constraints obtained by path integral integration of one of the conjugated momenta, i.e. 
$\pi_{f}=\pm \dot{f}e^{-\phi}$ for integrating out $\pi_{\bar{f}}$. 


Finally, accounting the for the complexified structure of the phase space, the $(f,\bar{f},\phi)$ variables of the 1D Schwarzian Model can be related to the $dS_{3}$ metric parameters by means of the line element expression 
\be{metric}
ds^{2}=d\phi^{2}-2e^{-\phi}df d\bar{f}\, . 
\ee

This series of relations deeply connects the Schwarzian quantum mechanics to the $dS_{3}/CFT_{2}$. In the next sections, we will see how TQG can also fit in this holographic picture. 

In Fig.1, we summarize the main correspondences among the models. 

\begin{figure}[t!!]
\begin{center}
\vspace{1cm}
\includegraphics[width=18cm,height=7cm,angle=0]{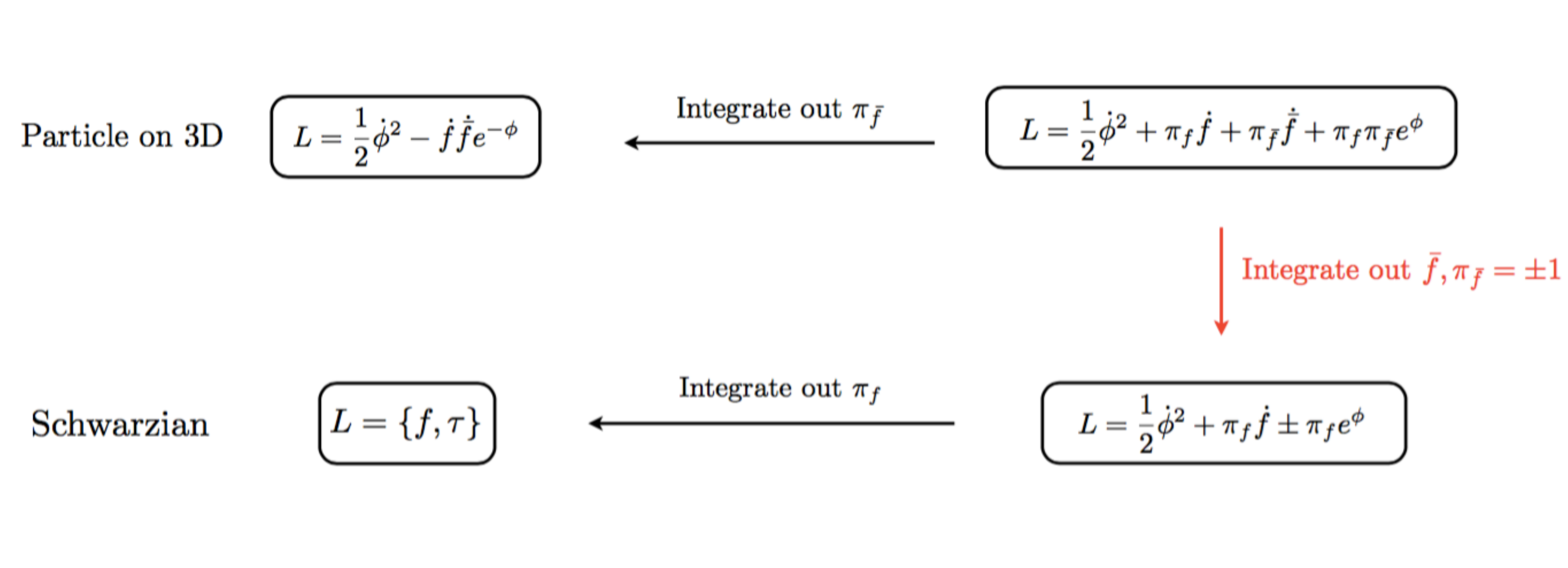}
\caption{We summarize a series of relations among the Schwarzian quantum mechanics with the particle motion in $dS_{3}$.  } \label{fig:1}
\end{center}
\end{figure}

\section{Solving the Schwarzian model from Conformal Bootstrap}

The 1D Schwarzian QM is related to a large $c$ limit of the 2D Virasoro CFT --- see Ref.~\cite{Mandal:2017thl}. The out-of-time order (OTO) function in the Schwarzian Model corresponds to the two-function of two bulk Liouville vertex operators. This correspondence allows to solve the Schwarzian Model as a limit of the 6j-symbols of the Virasoro CFT. 
A general expression for the 6j-symbol in the Virasoro CFT, constructed at finite $c$, and 
the relation with the monodromy of the conformal blocks were found in Ref.~\cite{Ponsot:1999uf}. The result in the large $c$ limit was recently found in Ref.~\cite{Mertens:2017mtv},  and can be cast in a compact form by means of
\be{jla}
\left\{\begin{array}{ccc} l_{1} & k_{2} & k_{s}
\ \\  l_{3} & k_{4} & k_{t} \ \\
\end{array} \right\} =\sqrt{\Gamma(l_{1}\pm ik_{2}\pm ik_{s})  \Gamma(l_{3}\pm ik_{2}\pm ik_{t}) \Gamma(l_{1}\pm ik_{4}\pm ik_{t})  \Gamma(l_{3}\pm ik_{4}\pm ik_{s})}
\ee
$$\times {\bf W}(k_{s},k_{t};l_{1}+i k_{4}, l_{1}-i k_{4}, l_{3} - i k_{2}, l_{3}+ i k_{2})\, , $$
where ${\bf W}(a,b,c,d,e,f)$ denotes the Wilson function, which in turn is composed by a linear combination of two generalized hypergeometric functions ${}_{4}F_{3}$ that reads
\be{xx}
 {\bf W}(\alpha,\beta;a,b,c,d)\equiv \frac{\Gamma(d-a)\, {}_{4}F_{3}\left[ \begin{array}{cccc} a + \imath \beta & a- \imath \beta  & \tilde{a} + \imath \alpha &\ \tilde{a} - \imath \alpha \\ a+b & a+c & 1+a-d &\end{array}\,; 1 \right] }{\Gamma(a+b) \Gamma(a+c) \Gamma(d\pm \imath \beta) \Gamma(\tilde{d}\pm \imath \alpha) }
+(a\leftrightarrow b) \,,
\ee
having defined 
\be{xy}
\tilde{d}=\frac{1}{2}(b+c+d-a)\,, \qquad \tilde{a}=\frac{1}{2}(a+b+c-d)\,,
\ee
and with
\be{xz}
{}_{4}F_{3}= \left[ \begin{array}{cccc} a_1 & a_2 & a_3 &\ a_4 \\ b_1 & b_2 & b_3 &\end{array}\,; z \right] \equiv \sum_{n=0}^{\infty} \frac{(a_1)_n \,(a_2)_n \,(a_3)_n \,(a_4)_n}{(b_1)_n \,(b_2)_n \,(b_3)_n } \frac{z^n}{n!} \,,    
\ee
where the $(a)_n$ coefficients are expressed in terms of the Euler $\Gamma$-functions, {\it i.e.}  
\be{dff}
(a)_n=\frac{\Gamma(a+n)}{\Gamma(a)}\,.
\ee
The notation $\Gamma(x\pm y\pm z)$ finally corresponds to taking the product of all the four sign combinations. 

\subsection{TO and OTO 4-point functions in Schwarzian quantum mechanics}

We summarize in this section how the out of time-ordered (OTO) 4-point functions can be divided from the knowledge of the time-ordered (TO) correlation functions. The latter ones are uncrossed planar diagrams while the former ones involve crossing of the legs. The diagrams associated to these correlations functions are connected by a crossing kernel that is expressed in terms of the R-matrix coefficients $R_{k_{s}k_{t}}$, as shown below.

The bi-local correlation operator is expressed by the formula 
\be{OOO}
O_{l}(\tau_{1},\tau_{2})=\left( \frac{\sqrt{f'(\tau_{1}) f'(\tau_{2})}}{\frac{\beta}{\pi}\sin \frac{\pi}{\beta}[f(\tau_{1})-f(\tau_{2})]} \right)^{2l}\, ,
\ee
and is invariant under SL$(2,\mathbb{R})$ transformations. This implies immediately that $O_{l}$ commutes with the Hamiltonian $H$ of the Schwarzian theory which is nothing but the quadratic Casimir operator of $\mathfrak{sl}(2,\mathbb{R})$ algebra:
\be{hol}
 [H,O_{l}(\tau_{1},\tau_{2}) ]=0\,.  
\ee    
Since the bi-local operators are diagonal in the energy-states basis, this implies that TO correlation functions $O_{l}(\tau_{1},\tau_{2})$ only depend on $\tau_2-\tau_1$.

At finite temperatures, the two-point function is defined by the single insertion of the bi-local operator in the functional integral $\mathcal{Z}=\int \mathcal{D} f \, \exp \{ -S[f] \}$, namely
\be{ozzl}
\langle O_l(\tau_1,\tau_2) \rangle = \frac{1}{\mathcal{Z}}\, \int \mathcal{D} f \, e^{-S[f]}\, O_l(\tau_1, \tau_2)\,.
\ee
At zero temperature, the two-point function of the Schwarzian theory was recovered in Ref.~\cite{zoto}, and as shown in \cite{Mertens:2017mtv}, the result can be generalized to finite temperature and expressed by a double integral over intermediate SL$(2,\mathbb{R})$ representations $k_1$ and $k_2$ by means of 
\be{oltolilo}
\langle O_l(\tau_1, \tau_2) \rangle = \int \prod_{i=1}^2 d\mu(k_i) \,\mathcal{A}_2(k_i,l,\tau_i)\,,
\ee
where $\mathcal{A}_2(k_i,l,\tau_i)$ are obtained in the large $c$ limit of previously recovered results, thanks to the relation between the Schwarzian theory and the 2D Virasoro CFT, and read
\be{A2}
\mathcal{A}_2(k_i,l,\tau_i)= e^{-(\tau_2-\tau_1)k_1^2 - (\beta -\tau_2+\tau_1) k_2^2} \, \frac{\Gamma(l\pm \imath k_1 \pm \imath k_2)}{\Gamma(2 l)}\,.
\ee

The 4-point operators $G_{l_{1}l_{2}}$ can be calculated from the bi-local operators $O_{l}(\tau_{1},\tau_{2})$ considering the expectation value 
\be{gaha}
G_{l_{1}l_{2}}(\tau_{1},\tau_{2},\tau_{3},\tau_{4})=\langle O_{l_{1}}(\tau_{1},\tau_{2}) O_{l_{2}}(\tau_{3},\tau_{4})\rangle \, . 
\ee
TO 4-point functions that are cyclically ordered via $\tau_1<\tau_2<\tau_3<\tau_4$ do not retain crossing of the legs of the bi-local operators involved. A triple integral expression over intermediate momenta can be immediately recovered, namely 
\be{TO4}
\langle O_{l_1} (\tau_1,\tau_2) O_{l_2} (\tau_3,\tau_4) \rangle_{\rm TO}= \int \prod_{i=1}^4 d\mu(k_i)\, \mathcal{A}_4 (k_i, l_i, \tau_i)\,.
\ee
In the momentum space, the expression of the $\mathcal{A}_4 (k_i, l_i, \tau_i)$ amplitudes for the TO 4-points functions was found in Ref.~\cite{Mertens:2017mtv}, taking into account the fact that bi-local operators commute with the Hamiltonian, as shown in Eq.~\eqref{hol}. This reads  
\be{A4TO}
\mathcal{A}_4 (k_i, l_i, \tau_i)= e^{-k_1^2 (\tau_2-\tau_1) - k_4^2 (\tau_4-\tau_3) - k_1^2 (\tau_2-\tau_1)- k_s^2 (\beta-\tau_2+\tau_3-\tau_4+\tau_1 )} [\gamma_{l_1}(k_1,k_2)]^2 \, [\gamma_{l_1}(k_1,k_2)]^2\,.
\ee

Here the Feynman rules in the momentum space have been applied, based on combinatoric algorithms that involve the propagators and the vertices. Specifically, the propagator between $\tau_1$ and $\tau_2$ that is labelled by $k$ is calculated to be $\exp -k^2 (\tau_2-\tau_1)$, while the vertex factor intertwining the momenta $l$, $k_1$ and $k_2$ is found to be
\be{glk1lk2}
\gamma_l(k_1,k_2)= \sqrt{\frac{\Gamma(l\pm\imath k_1\pm \imath k_2)}{\Gamma(2l)}}\,.
\ee
As shown in Ref.~\cite{Mertens:2017mtv}, the same results for 4-point function can be also recovered from relation between the Schwarzian Model and the 2D CFT. 

Notice furthermore that the amplitude in Eq.~\eqref{A4TO} factorizes into the product of two 2-point amplitudes
\be{fac}
\mathcal{A}_4 (k_i, l_i, \tau_i)= e^{\beta k_s^2} \mathcal{A}_2 (k_1, k_s, l_1, \tau_{2}-\tau_{1})\,\mathcal{A}_2 (k_4, k_s, l_2, \tau_{4}- \tau_{3})\,.
\ee

The OTO 4-point function $\langle O_{l_1} (\tau_1,\tau_2) O_{l_2} (\tau_3,\tau_4) \rangle_{\rm OTO}$ was calculated in Ref.~\cite{Mertens:2017mtv}, deploying an analytic continuation, starting from the TO correlation functions with time ordering $\tau_1<\tau_3<\tau_2<\tau_4$. Since in this latter case the legs of the bi-local operators cross, the resulting TO 4-point function will differ from the analytic continuation of the 4-point function recovered without crossings. This entails the expression 
\be{OTO4}
\langle O_{l_1} (\tau_1,\tau_2) O_{l_2} (\tau_3,\tau_4) \rangle_{\rm OTO}= \int \prod_{i=1}^4 d\mu(k_i)\, \mathcal{A}^{\rm OTO}_4 (k_i, l_i, \tau_i)\,,
\ee
in which now the TO and OTO amplitudes are related through the R-matrix coefficients $R_{k_{s}k_{t}}$ by means of
\begin{eqnarray}\label{Gllkl}
\mathcal{A}^{\rm OTO}_4 (k_i, l_i, \tau_i)=&&
e^{- k_1^2 (\tau_3-\tau_1) - k_t^2 (\tau_3-\tau_2)  - k_4^2 (\tau_4-\tau_2) - k_s^2 (\beta-\tau_4+\tau_1)  } \\ 
&&\times \gamma_{l_1}(k_1,k_s)\, \gamma_{l_2}(k_1,k_4)\, \gamma_{l_1}(k_4,k_t)\, \gamma_{l_2}(k_t,k_1)\,
\times R_{k_{s}k_{t}}\left[\begin{array}{ccc} k_{4} & l_{2} 
\ \\  k_{1} & l_{1}  \end{array} \right]\,. \nonumber
\end{eqnarray}

The R-matrix coefficients depend on six numbers, namely $k_{1}$, $k_4$, $k_s$, $k_t$, $l_{1}$ and $l_{2}$, labelling the spin of the corresponding sextuplet or representations of SL$(2,\mathbb{R})$. 

In general, the $R$-matrix satisfies the unitarity property 
\be{unitarity}
\int d\mu(k) R_{k_{s}k}R^{\dagger}_{k k_{t}}=\frac{1}{\rho(k_{s})}\delta(k_{s}-k_{t}), \qquad {\rm with} \qquad  \rho(k)=2k \sinh(2\pi k)\, ,
\ee
while its explicit form of can be recovered considering the 2D CFT side of the Schwarzian Model. Intriguingly, it turns out that the 2D kernel can be expressed as a quantum 6j-symbol of the non-compact quantum group U$_{q}(\mathfrak{sl}(2,\mathbb{R}))$, as shown in Eq.~(\ref{jla}). 

\subsection{Correspondence dictionary}
\noindent 
A dictionary between the Schwarzian Model and the $CFT_{2}$ theory can be individuated, which can be useful for calculations. This entails the correspondence
\be{Inserti}
{\rm Schwarzian:}\,\,\, O_{l}(\tau_{1},\tau_{2}) \qquad \longleftrightarrow \qquad {\rm Liouvile\,\, CFT:}\,\,\, V_{l}=e^{2l\phi(\tau_{1},\tau_{2})}\, ,
\ee
which leads to the relation between the Liouville and the Schwarzian fields
\be{jjjal}
e^{\phi_{Cl}(u,v)}=\frac{\sqrt{f'(u)f'(v)}}{\frac{\beta}{\pi}\sin \frac{\pi}{\beta}[f(u)-f(v)]}\, . 
\ee
In other words, the bi-local operator in the 1D Schwarzian Model corresponds to the insertion of the $V_{l}$ vertex between two ZZ-branes. \\

As a consequence, the 4-point function corresponds, in 2D Liouville CFT, to the 2-point function of primary operators between two ZZ-branes $|ZZ\rangle$, namely 
\be{ZZZ}
G_{l_{1}l_{2}}=\langle ZZ| V_{l_{1}}(z_{1},\bar{z}_{1}) V_{l_{2}}(z_{2},\bar{z}_{2})|ZZ\rangle \, , 
\ee
provided the identifications 
$z_{1}\rightarrow \tau_{1}$, 
$\bar{z}_{1}\rightarrow \tau_{2}$,
$z_{2}\rightarrow \tau_{3}$
and $\bar{z}_{2}\rightarrow \tau_{4}$. When TO operators are taken into account, the sites $\{z_1,\bar{z}_1\}$ and $\{z_2,\bar{z}_2\}$ are taken to be timelike separated, which ensures that their past light-cones do not intersect. When the two bulk operators are spacelike separated, following 2D light-cone directions that originate from each vertex, the legs of the bi-local operators show a crossing. Then TO and OTO correlation functions then turn out to be related by the CFT monodromy matrix. This latter links the timelike separated and the spacelike separated 2-point functions.\\

The R-matrix of the 2D Virasoro conformal blocks equals 
\be{hk}
R_{\alpha_{s}\alpha_{t}}\left[\begin{array}{ccc} \alpha_{3}& \alpha_{2} 
\ \\  \alpha_{4} & \alpha_{1} \ \\
\end{array} \right]=e^{2\pi \imath (\Delta_{2}+\Delta_{4}-\Delta_{s}-\Delta_{t})}F_{\alpha_{s}\alpha_{t}}\left[\begin{array}{ccc} \alpha_{3}& \alpha_{2} 
\ \\  \alpha_{4} & \alpha_{1} \ \\
\end{array}\right]\,,
\ee
where $F_{\alpha_{s}\alpha_{t}}$ is expressed in terms of the quantum 6j-symbol by means of
\be{hjk}
F_{\alpha_{s}\alpha_{t}}=|S_{b}(2\alpha_{t})S_{b}(2\alpha_{s})|\sqrt{\frac{C(\alpha_{4},\alpha_{t},\alpha_{1})C(\bar{\alpha}_{t},\alpha_{3},\alpha_{2})}{C(\alpha_{4},\alpha_{3},\alpha_{s})C(\bar{\alpha}_{s},\alpha_{2},\alpha_{1})}}
\left\{\begin{array}{ccc} \alpha_{1} & \alpha_{2} & \alpha_{s}
\ \\  \alpha_{3} & \alpha_{4} & \alpha_{t} \ \\
\end{array} \right\}\,.
\ee
In Eq.~\eqref{hjk}, $S_{b}(x)$ denotes the double Sine function, defined by
\be{ssss}
S_{b}(x) \equiv \frac{\Gamma_b(x)}{\Gamma_b(Q-x)}
\,,
\ee
with $Q=b+b^{-1}$ and $\Gamma_2(z| a_1, a_2)$ the Barnes double gamma function\footnote{In general, the multiple gamma functions are defined by the properties 
\begin{eqnarray} \label{Bdbf}
&\Gamma_0(z|)=\frac{1}{z}\,, \qquad \Gamma_1(z|a)=\frac{a^{a^{-1 z -\frac{1}{2}}}}{\sqrt{2 \pi}} \Gamma(a^{-1}z)\,, \nonumber \\ 
&\Gamma_N(z|a_1, \dots, a_N) = \Gamma_{N-1}(z|a_1, \dots, a_{N-1})\, \Gamma_{N}(z+a_N|a_1, \dots, a_{N}) \,. \nonumber
\end{eqnarray}
}. 
It is relevant for what follows that in the $b\rightarrow{0}$ limit one finds 
\be{gammaslimits}
\Gamma_b(bx) \rightarrow (2\pi b^3)^{\frac{1}{2} (x-\frac{1}{2})} \Gamma(x)\,, \qquad \Gamma_b(Q-bx) \rightarrow (2\pi b)^{-\frac{1}{2} (x-\frac{1}{2})} \,,
\ee
implying the limit $$S_{b}(x) \rightarrow (2\pi b^3)^{x-\frac{1}{2}} \Gamma(x)\,.$$
The $C(\alpha_{3},\alpha_{2},\alpha_{1})$ coefficients entering Eq.~\eqref{hjk} denote the DOZZ 3-point functions \cite{Dorn:1994xn,Zamolodchikov:1995aa,Teschner:1995yf}, while the momenta are rescaled according to $\alpha_{i}=b\, l_{i}$.

It is remarkable that in Eq.~(\ref{hk}) the q-deformed version of the 6j-symbol appears, with $q=e^{\imath \hbar}=e^{\imath \pi b^{2}}$. In view of the relation with 3D gravity that we will discuss in the next session, we remind that the q-deformation of the 6j-symbol, which is relevant in 3D Topological Quantum Gravity, deploys $q=e^{\imath \hbar}=e^{\imath {\sqrt{\Lambda}}/{2}}$, with $\Lambda$ cosmological constant.\\ 
  
Finally, we remark that for the application to the Schwarzian theory, we are interested in the $b\rightarrow 0$ limit, then correspondent on the gravitational side to zero cosmological constant limit, of the quantum 6j-symbols. This limit entails 
\be{ak}
\left\{\begin{array}{ccc}l_{1} & k_{2} & k_{s}
\ \\  l_{3} & k_{4} & k_{t} \ \\
\end{array} \right\}=\lim_{b\rightarrow 0} 2\pi b^{3} \left\{\begin{array}{ccc} \alpha_{1} & \alpha_{2} & \alpha_{s}
\ \\  \alpha_{3} & \alpha_{4} & \alpha_{t} \ \\
\end{array} \right\}_{b}\, . 
\ee  
This finally leads to the expression of the Schwarzian R-matrix coefficients 
\be{hak}
R_{k_{s}k_{t}}\left[\begin{array}{ccc} k_{4} & l_{2} 
\ \\  k_{1} & l_{1} \ \\
\end{array} \right]=\left\{\begin{array}{ccc} l_{1} & k_{1} & k_{s}
\ \\  l_{2} & k_{4} & k_{t} \ \\
\end{array} \right\}\, .
\ee
  
\section{Gravitational scattering in $dS_{3}/CFT_{2}$}  
  
In the previous section we reviewed the relation between the the Schwarzian Model and the $CFT_2$ theory and the fundamental result recovered in Ref.~\cite{Mertens:2017mtv} that the behavior of the OTO correlation functions is dictated by the knowledge of the TO correlation functions, once the R-matrix 6j-symbols are known. The R-matrix is explicitly unitary and describes the gravitational bulk scattering amplitude in the momentum space. In this section, we come back to the analysis of the theory on the $dS_3$ side, and shed light on further correspondences that be recovered with  2+1D non-commutative space-times in the limit of vanishing cosmological constant. 

\subsection{Particle scattering in $dS_3$}

We start the analysis considering a BTZ black hole in $dS_{3}$, the metric of which reads 
\be{dss}
ds^{2}=\left(\frac{r^{2}+R^{2}}{l^{2}}\right)dt^{2}- \left(\frac{r^{2}+R^{2}}{l^{2}}\right)^{-1}dr^{2}+r^{2}d\phi^{2}\,, \quad {\rm with} \quad R^{2}=8Ml^{2}\,, 
\ee
where $l$ is the dS radius and $R$ the Schwarzschild radius. This suggests that the dynamics of every 3D Schwarzschild De Sitter can be completely captured by the CFT on the boundary of $dS$. As a remarkable consequence, every particle scattering on the dS geometry must have an R-matrix, which is equivalent to the one recovered on the CFT and Schwarzian sides. A Penrose diagram of the scattering process is shown in Fig.~1, in which the conformal coordinates are deployed 
\be{njk}
ds^{2}=-\frac{1}{\cos^{2}T}(dT^{2}+d\Omega_{2}^{2})\, ,
\ee
with $-\pi/2<T<\pi/2$. 

\begin{figure}[t!!]
\begin{center}
\vspace{1cm}
\includegraphics[width=5cm,height=5cm,angle=0]{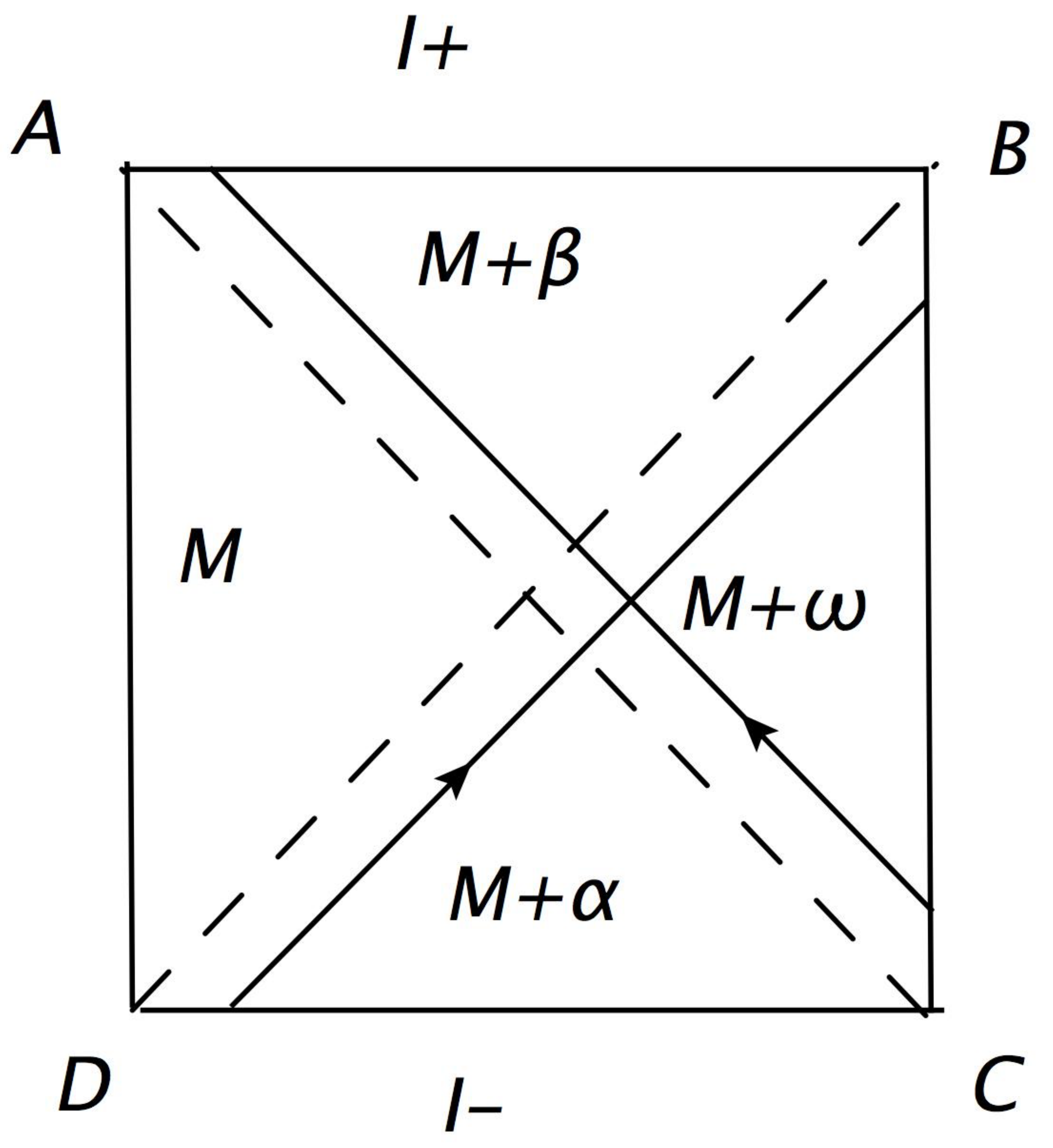}
\caption{Scattering of two particles close to the horizon of a BTZ black hole in $dS_{3}$,
in the Penrose diagram representation. 
The AD and BC sides are timelike lines. Every point in the interior represents
an $S^{1}$ circle, while a horizontal slice is a $S^{2}$ sphere. 
The dashed lines are the past and future horizons before the scattering. 
The entire space-time, and the scattering process, cannot be observed by one only observer. 
One observer $O^{-}$ can cover the BDC triangle, 
while an $O^{+}$ can access to ABC. } \label{fig:1}
\end{center}
\end{figure}

The scattering of two particles on the $dS_3$ background can be addressed following the description deployed in Refs.~\cite{Jackson:2014nla,Mertens:2017mtv}. We denote with $M$ the mass of the black hole, and label with the ``$\phantom{a}_{(1)}$'' and ``$\phantom{a}_{(2)}$'' the two particles that are respectively departing and arriving at the dS boundary, and that collide close to the horizon. The scattering process creates a future black hole region with a shifted mass, $M+\beta$. The value of $\beta$ is determined by the collision energy, which is in turn dependent on $M$, the energies $\alpha$ and $\omega$ of the particle modes, and the time difference between ``$\phantom{a}_{(1)}$'' and ``$\phantom{a}_{(2)}$''. The mass parameters entering in the scattering problem are 
\begin{eqnarray} \label{MMMMA}
M_{1}=M\,, \qquad  M_{2}=m_{1}\,,\qquad M_{\alpha}=M+\alpha\, , \nonumber \\
M_{3}=M+\omega,\, \qquad M_{4}=m_{2},\, \qquad  M_{\beta}=M+\beta\, . 
\end{eqnarray}
Consistently with this notation, we may define the $\phi^{(1)}$ infalling modes and the $\phi^{(2)}$ outgoing modes. The scattering process then formally corresponds to the expression 
\be{hakal}
\phi_{\omega-\alpha}^{(2)}\phi_{\alpha}^{(1)}=\sum_{\beta}R_{\alpha \beta}\phi_{\omega-\beta}^{(1)}\phi_{\beta}^{(2)}\, , 
\ee
where $R_{\alpha\beta}$ is the unitary scattering operator which shifts the location of one operator by an amount proportional to the energy of the other one.\\

We may then consider a 2D CFT endowed with large central charge $c$ and a holographic dual that is weakly coupled. The theory has dense spectrum of primary states $|M\rangle$, with the Cardy formula governing their asymptotic level density. Resorting to the dual gravity description, the heavy states $|M\rangle$ endowed with energies $l \, M \!>\! \frac{c}{12}$ shall describe the BTZ black holes accounted in Eq.~\eqref{dss}. 

Particles can be introduced in this framework as perturbations escaping from near the horizon of the black hole, or infalling into the latter. Thus a generic state $|\Psi\rangle$ can be obtained in the Heisenberg picture by acting with a local CFT operator $\hat{O}$ on $|M\rangle$. A specific operator  that creates a perturbation/particle ``$\phantom{a}_{(1)}$'' is denoted as $\hat{O}^{(1)}$. Its action can be expanded by means of 
\be{AM}
|\Psi\rangle=\hat{O}^{(1)}|M\rangle=\sum_{\alpha}\hat{O}_{\alpha}^{(1)}(t_{0})|M\rangle\,,
\ee
in which the components capture the increase of the primary states' energy, namely 
\be{increasing}
\hat{O}^{(1)}_{\alpha}|M\rangle=e^{-i\alpha t_{1}}f_{\alpha}^{(1)}|M+\alpha \rangle+{\rm descendants}\, ,
\ee
where ``$f_{\alpha}^{(1)}$'' denote the OPE coefficients. Descendants are given by product of Left and Right Virasoro generators that act on the states $|M+\alpha\rangle$. The components $\hat{O}^{(1)}_{\alpha}$ are associated to the description of the partial wave for which the bulk particle ``$\phantom{a}_{(1)}$'' has energy $\alpha$. 
Along the same line, we introduce the second particle operator, which acts at an earlier time $t_{2}<t_{1}$ according to 
\be{BAA}
\hat{B}_{\omega-\alpha}(t_{2})\hat{A}_{\alpha}(t_{1})|M\rangle
=e^{-i\omega t_{2}-i\alpha(t_{2}-t_{1})} f_{\omega-\alpha}^{(2)}f_{\alpha}^{(1)}|M +\omega \rangle+{\rm descendents}\,,
\ee
where $\omega$ is the total energy of the bulk particles.\\ 

On the CFT side, Eq.~(\ref{hakal}) recasts as 
\be{KLa}
\hat{B}_{\omega-\alpha}(t_{2})\hat{A}_{\alpha}(t_{1})|M \rangle=\sum_{\beta}R_{\alpha \beta}\hat{A}_{\omega-\beta}(t_{1})\hat{B}_{\beta}(t_{2})|M\rangle \, . 
\ee
The scattering matrix that is recovered from the side of the conformal bootstrap is the one reported in Eq.~(\ref{hk}). Within the semi-classical scattering regime, the R-matrix can be then related to the S-matrix, which in the eikonal form reads 
\be{RS}
R_{\alpha\beta}={\rm exp}\left(\frac{i}{\hbar}S_{\alpha\beta}\right)\, .
\ee
In the limit of small quantum corrections, namely $\hbar\rightarrow 0$, up to trivial phases one obtains   
\be{kajka}
S_{\alpha\beta}={\rm Vol}\left(T\left[\begin{array}{ccc} 1 & 2 & \alpha
\ \\  3 & 4 & \beta \ \\
\end{array} \right] \right)\, .
\ee
The S-matrix corresponds to the volume of a generic tetrahedron \cite{T1,T2,T3}, with angles, at the six edges, controlled by the mass parameters in Eq.\eqref{MMMMA} through $l_{i}/2\pi=\sqrt{8M_{i}}$. 

\subsection{Relating the Schwarzian Model with noncommutative spacetimes}

We close this section emphasizing a further correspondence that can be recovered from the side of the $dS_3$ theory, and that traces back to a longstanding analysis of topological scattering, as addressed in Refs.~\cite{Lo:1993hp, Wilczek:1989kn}, which are relevant in lower dimensional quantum gravity Refs.~\cite{Bais:1998yn,Bais:2002ye,Marciano:2010jm, Marciano:2010gq}. In particular, using a Chern-Simons formulation 2+1D gravity without cosmological constant, it was shown in \cite{Bais:2002ye} that gravitational interactions deform in flat spacetime the Poincar\'e symmetry up to a quantum group of symmetries. This latter turns out to be the quantum double of the universal cover of the 2+1D Lorentz group, namely a Lorentz double. The R-matrix of the Lorentz double can be then deployed to address the scattering of gravitating particles endowed with spin, and the relevant results obtained reproduce previous findings by 't Hooft \cite{S1}, Deser and Jackiw \cite{S2}  and de Sousa Gerbert and Jackiw \cite{S3}. Moving from the quantum deformation of the isometry algebras of the 2+1D $AdS$, $dS$ and Minkowski spacetimes, 
this result was extended in Ref.~\cite{Ballesteros:2014kaa} to the correspondent Drinfel'd double structure on the isometry algebras that are motivated by their role in (2+1)-gravity, in order to encode the cosmological constant $\Lambda$ as the deformation parameter, and obtain twisted versions of both space-like and time-like $\kappa$-$AdS$ and $\kappa$-$dS$ in quantum algebras.

It is then relevant that in \cite{Ballesteros:2014kaa} the $\Lambda \rightarrow 0$ limit leads to a twisted quantum Poincar\'e algebra, with related non-commutative spacetime that is a nonlinear $\Lambda$-deformation of $\kappa$-Minkowski plus additional contribution due to the twist. This clearly corresponds to the $b\rightarrow 0$ limit shown in Eq.~\eqref{ak}, through which we derive the Schwarzian Model. This then amounts to the identification of a new correspondence that in the $b \rightarrow 0$, or equivalently  $\Lambda \rightarrow 0$, relate the Schwarzian Model to the non-commutative $\kappa$-$dS$ spacetime. A similar result can be immediately derived on the $AdS$ side, establishing a correspondence between the Schwarzian Model and the non-commutative $\kappa$-$AdS$ spacetime. For this latter case, it was shown in \cite{Ballesteros:2014kaa} that the deformation that is recovered can be related to two copies of the standard  quantum deformation of the Lorentz group in 3D, also referred to as Drinfel'd-Jimbo deformation.

\section{3D quantum gravity, Kac-moody instantons and holonomies}

In this section, we further refine our analysis on the $dS_{3}$ side, and elaborate on the interesting results obtained in Ref.~\cite{Banados:1998tb}. 

We start performing a Wick rotation of Eq.~\eqref{jjj}, so to cast the Euclidean de Sitter background as
\be{EdS}
ds_{E}^{2}=\left(1-\frac{r^{2}}{l^{2}}\right) dt_{E}^{2}+\left(1-\frac{r^{2}}{l^{2}}\right)^{-1}dr^{2}+r^{2}d\theta^{2}\, , 
\ee
where $\beta=2\pi l$ is the period of the time coordinate $t_{E}$, with $0 \leq t_{E}<\beta$.

Considering the change of coordinates $r=l\cos \rho$, we obtain the $S_{3}$ sphere metric 
\be{threesphere}
ds_{E}^{2}=\sin^{2}\rho \, dt_{E}^{2}+l^{2}\, d\rho^{2}+l^{2}\cos^{2}\rho \, d\theta^{2}\, . 
\ee
The Euclidean sector can be described by a double Chern-Simons gauge theory, with gauge group SU$_{+}(2)\times$ SU$_{-}(2)$. Each SU$_{\pm}(2)$ chiral sector is characterized by the gauge connection 
\be{AAL}
A^{a}_{\pm}=\omega^{a}\pm \frac{1}{l}e^{a}\, ,
\ee
in which $\omega^{a}$ and $e^{a}$ denote respectively the spin connection and triad.\\

We can explicitly find the components of the connection  introducing a generalization on the Euclidean $dS_{3}$ background in terms of two real parameters, $\gamma$ and $\beta$. The connection components then read  
\be{jak}
A_{\pm}^{1}=- \gamma\sin \rho\, \left( d\theta \, \mp \, \frac{\beta}{l}dx^{0} \, \right)\, ,
\ee
\be{jka}
A_{\pm}^{2}=\pm \, d\rho\, ,
\ee
\be{jaka}
A_{\pm}^{3}=\gamma\cos \rho \left(d\theta \, \mp \, \frac{\beta}{l}dx^{0}\right)\, , 
\ee
where $\gamma=1$ corresponds to the $dS_3$ background. The requirement that no conical singularities or holonomies are present at the horizon ($\rho=0$) fixes the ratio between the two parameters to be $\beta=2\pi l \gamma^{-1}$.\\

In what follows we focus only on the positive-chirality sector of the theory, and neglect the specular, negative-chirality, mirror sector. Furthermore, we consider that the horizon at $\rho=0$ is associated to a boundary condition on the connection, by means of
\be{lal}
A_{0}^{a}|_{\rho=0}=-2\pi\, \delta_{3}^{a}\, . 
\ee
The Euclidean action that fits the boundary conditions then reads
\begin{eqnarray} \label{Reads}
I_{E}[A,\beta]=&& \phantom{a} \frac{k}{4\pi}\int_{M}\epsilon^{kl}{\rm Tr}(i A_{k} \partial_{0} A_{l}-A_{0}F_{kl})d^{2}xdx^{0} \nonumber\\
&&-\frac{k\beta}{4\pi l}\int_{\rho=\pi/2}{\rm Tr}(A_{\phi})^{2}d\theta dx^{0}-\frac{k}{2}\int_{\rho=0}A_{\phi}^{(3)}d\theta dx^{0}\, , 
\end{eqnarray}
where the level of the Chern-Simons theory reads $k=l/4G$. 

The first term of Eq.~\eqref{Reads} is nothing but the Euclidean Chern-Simons action on a manifold with topology $M=\Sigma \times S_{1}$, where $S_{1}$ denotes the compactified time direction with period $\beta^{-1}$. \\

It is remarkable to note that, performing the integral of the Euclidean Chern-Simons action, a hidden Kac-Moody algebra at level $k$ can be found. To show this result, we first consider the correspondent partition function, which reads  
\be{jakla}
Z_{A}(\beta)=\int \mathcal{D}A e^{-I_{E}[A;\beta]}\, .
\ee
The semiclassical limit of Eq.~\eqref{jakla} is achieved through the saddle point approximation, considering the gauge components Eqs.~(\ref{jak})-
(\ref{jaka}), namely 
\be{AJAKL}
Z_{A}(\beta)=\frac{1}{N}\, \sum_{\gamma}{\rm Exp}\left(-\frac{1}{16G}\beta \gamma^{2}+\frac{1}{4G}\pi \gamma l\right)\, , 
\ee
where $N$ denotes a normalization factor. The partition function is recovered as a sum performed over $\gamma$ at $\beta$ fixed. The saddle point approximation then yields the $\gamma \beta = 2\pi l$, corresponding to the classical value of $\gamma$ that avoids conical singularities at the horizon. It is worth to emphasize that the expression for $Z(A)$ in Eq.~\eqref{AJAKL} is reminiscent of the Regge discretization of the action of gravity in 2+1D, when the configuration variables, the lengths of the edges of the simplicial triangulated manifold, are assumed to be proportional to $\gamma l$, and the cosmological constant term is reabsorbed in the normalization factor. From the geometrical point of view, $\gamma$ can be also seen to parametrize the deficit angle of a conical singularity located at $\rho=\pi/2$ ($r=0$). From Eq.(\ref{AJAKL}), we can associate for each $\gamma$-state an energy $E_{\gamma}=\gamma^{2}/16G$ and a degeneracy $\rho(\gamma)={\rm exp}(\pi \gamma l/(4G))$.

It is worth to note that the $\gamma$-states satisfy the Euclidean EoM of the theory, since are associated to the $\gamma$-families of $dS$ solutions. This implies that these states correspond to instantonic solutions, and thus may be interpreted as quantum mechanical tunnelings. Since $\gamma\neq 1$ amounts to pick conic singularities in the black hole horizon, the instantonic solutions can be connected to the gravitational tunneling from the interior to the exterior of the black hole horizon. In other words, the instantons puncture the black hole horizon. On the other hand, one should expect that the sum would not be supported over a continuum set of values, and that, somehow, the $\gamma$'s should be discretized and related to a topological winding number. Elaborating on the asymptotic limit, we will see that this is exactly what happens. 

It is also worth to note that for $\gamma=1$ (de Sitter solution) the total entropy (considering both the chirality sectors) turns out to be
\be{SSS}
S=S_{l}+S_{r}=\frac{\pi l}{2G}\, ,
\ee
which is nothing but the Gibson-Hawking value for the de Sitter background. \\

The model reduces to the WZW theory at the boundary, namely
\be{ICWZW}
I_{\rm CWZW}[g,\beta]=-\frac{\imath k}{4\pi}\int {\rm Tr}(\partial_{\phi}g^{-1}\partial_{\tau} g ) d\theta d\tau-\frac{\imath k}{12 \pi}\int_{M}{\rm Tr}(g^{-1}d g)^{3} d^{2}x d\tau
\ee
$$-\frac{k\beta}{4\pi l}\int {\rm Tr}(g^{-1}\partial_{\theta} g)^{2} d\theta d\tau -\frac{k}{2}\int (g^{-1}\partial_{\phi})^{(3)}d\theta d\tau\, , $$
where $g$ is an element of the group, which turns out to be 
not only the SU$(2)$ Lie group, but also a SU$(2)$ Kac-Moody. \\

The asymptotic gauge field can be expanded as 
\be{Aphi}
A_{\phi}=g^{-1}\partial_{\phi}g=\frac{2}{k}\sum_{n=-\infty}^{\infty}T_{n}^{a}\, e^{in \phi}\, , 
\ee
where $T_{n}^{a}$ satisfies the $SU(2)$ Kac-Moody algebra, namely  
\be{haklasjk}
[T_{n}^{a},T_{m}^{b}]=\imath\, \epsilon_{\ \ c}^{ab}\, T_{n+m}^{c}+n\frac{k}{2}\, \delta^{ab}\, \delta_{n+m,0}\, \, . 
\ee
This leads immediately to the partition function 
\be{part}
Z_{A}(\beta)=\sum_{2s=0}^{k}{\rm Tr}_{s} \left[ e^{ - \frac{\beta}{l}L_{0} +2\pi \, T_{0}^{3} }\right] \, , 
\ee
where 
\be{Lz}
L_{0}=\frac{1}{k+Q}\sum_{n=-\infty}^{n=\infty}:T_{-n}^{a}T_{n}^{b}:\delta_{ab}\, .
\ee
In Eq.~\eqref{Lz}, $L_{0}$ denotes the Virasoro generator, while $Q$ stands for the second Casimir operator in the adjoint representation of the group. 

We may now consider the semi-classical limit, namely $k\!>\!\!>\!Q$, in which it is possible to compute the trace on the spin-representations. Using the result of the ${\rm Tr}[q^{L_{0}}e^{\imath \, \theta \, T_{0}^{3}}]$ given in Ref.~\cite{12}, after some tedious but straightforward algebraic passages we recover
\be{jkl}
Z_{A}(\beta)=\sum_{2s}^{k}d^{-1} q^{\frac{s(s+1)}{k}}\frac{\sinh[2\pi(s+\frac{1}{2})]}{\sinh \pi}\sum_{n=-\infty}^{+\infty}
q^{k n^{2}+(2s+1)n}e^{2\pi k n}\, , 
\ee
\be{sjsks}
q=e^{-\beta/l}\,, \qquad  d=\prod_{n=1}^{\infty}(1-q^{n})(1-q^{n}e^{\imath \theta})(1-q^{n}e^{-\imath \theta})\, , 
\ee
where $\theta=-2\pi \imath $ in our specific case. 

In the $k\rightarrow \infty$ limit, the saddle point condition is obtained
\be{jaajaj}
n=\frac{\gamma}{2}-\frac{1}{2k}-\frac{s}{k}\, . 
\ee
Since both $n$ and $s$ are discrete, this means that also $\gamma$ is quantized. This result also provides an explicit 
example of the conjecture, proposed in a series of previous papers, that the partition function of a black hole may be interpreted as a superposition of conical naked singularities --- see e.g. Refs.~\cite{Addazi:2015gna,Addazi:2016cad,Addazi:2015hpa,Addazi:2015cho}. It is also worth to note that in Refs.~\cite{Addazi:2015gna,Addazi:2016cad,Addazi:2015hpa,Addazi:2015cho} a relation between conical singularities and chaotization of infalling information was pointed out. Indeed, the chaotization phenomenon can be recast from the prospective of the eikonal scattering, confirming the conjecture of Refs.~\cite{Addazi:2015gna,Addazi:2016cad,Addazi:2015hpa,Addazi:2015cho}.\\
  
Eq.~\eqref{jaajaj} is a very important result, since connects the gravitational instantons in the semiclassical limit 
to the spin representations of $SU(2)$, which in turn are connected to 3D TQG holonomies in the non-perturbative regime. 

In particular, in the limit of $k\rightarrow \infty$, we must have that $n<1$ and then that $n$ can be neglected, giving rise to the a discrete tower of conical instantons, namely  
\be{gaa}
\gamma=\frac{2s+1}{k}\, ,
\ee
with a corresponding partition function 
\be{kakaak}
Z_{A}[\beta]=\sum_{2s=0}^{k}e^{2\pi s}e^{-\beta s(s+1)/lk}\, .
\ee
The quantized energy levels and the degeneracy factor appears in the 
partition function:
\be{ajkaa}
E_{s}=\frac{s(s+1)}{lk}\,, \qquad \rho(s)=e^{2\pi s}\, . 
\ee
In particular, the degeneracy factor of the states leads
directly to the Gibson-Hawking entropy, which casts 
\be{SS}
S=2\pi k=\frac{2\pi l}{4G}\, . 
\ee

It is worth to note that this counting of states scales exactly as the area of a black hole in topological quantum gravity, which is recovered from counting the holonomy's punctures on the black hole's horizon. 

\section{Conclusions and remarks}

Within the context of $dS_{3}/CFT_{2}$, we have shown that a unification between TQG and string theory can be recovered, as in the case of $AdS_{3}/CFT_{2}$ \cite{Turiaci:2017zwd,Mertens:2017mtv}. We discovered an intriguing pattern of correspondences among 1D SYK, Schwarzian quantum mechanics, 3D double Chern-Simons theory, Liouvile $CFT_{2}$ and the eikonal scattering of particles on the surface of a black hole. In particular, the main building block of the quantum amplitudes for the thepreis considered here, the 6j-symbol, arises from every sides of $dS_{3}/CFT_{2}$ correspondence, leading to the connection between string theory and TQG. This seems to enforce the recent conjectures of unification among topological M-theory and TQG proposed in Refs.\cite{Addazi:2018cyn,Addazi:2017qwt}. \\

On the side of $dS_{3}$, we elaborated on Ref.~\cite{Banados:1998tb} and showed that in the semiclassical limit the partition function can be decomposed as an infinite sum of partition functions of conical euclidean singularities. These latter in turn are nothing but instantons of the double Chern-Simons theory. This observation provides an explicit example of the conjecture proposed in Refs.~\cite{Addazi:2015gna,Addazi:2016cad,Addazi:2015hpa,Addazi:2015cho}. In the semiclassical regime, the partition function of a black hole can be related to an infinite sum of partition functions of conical singularities. Another side of this conjecture has been then proved in this paper, since we suggested that the presence of conical singularities can be related to the chaotization of infalling information \cite{Addazi:2015gna,Addazi:2016cad,Addazi:2015hpa,Addazi:2015cho}. This proposal then finds within this context a paradigmatic example. The eikonal approach for describing two particle collisions close to the black hole horizon in $dS_{3}$ leads to the butterfly effect as a back-reaction phenomenon, as in the $AdS_{3}$ case studied in Ref.~\cite{Shenker:2013pqa}. \\

It is also remarkable that a correspondence can be found between the Schwarzian Model and non-commutative spacetimes too. We took into account previous results recovered in Ref.~\cite{Bais:2002ye}, which concern the emergence of a Lorentz double with R-matrix involved in the description of the scattering of gravitating particles, and in Ref.~\cite{Ballesteros:2014kaa}, which concern the recovery of twisted non-commutative spacetimes in the $\Lambda \rightarrow 0$ of deformed Drinfel'd double structures for $AdS$ and $dS$. Thanks to these findings, we recognized that the correspondence can be immediately extended in order to encode the Schwarzian Model in the $\Lambda \rightarrow 0$ limit.\\

We wish finally to summarize the very interesting set of properties retained by the instantons we found here: 
\begin{itemize}  
\item
Instantonic solutions are connected by a Kac-Moody global symmetry. A similar result, accounting for gauge and gravitational instantons on a $S_{2}\times S_{2}$ background,  was found in Ref.~\cite{Addazi:2017xur}.

\item
The instantonic levels are proportional to the spin representations of $SU(2)$. This paves a new way to obtain the semiclassical limit of TQG in $dS_{3}$. 

\item 
The instantonic solutions are corresponding to the BH entropy, which is obtained from the degeneracy factor of the instantonic states in the partition function. This provides a powerful example of a connection between the Kac-Moodions and the black hole entropy, as conjectured in Ref.~\cite{Addazi:2017xur}.

\end{itemize}  

These properties lead to an interesting interpretation of Kac-Moodions as gravitational tunneling solutions that puncture the BTZ black hole horizon. In other words, the picture we present  is the semi-classical analogous of the TQG holonomy punctures on the BH horizon. As a matter of fact, the two approaches match in the counting of the BH entropy states. It is also worth noting that the gravitational tunneling may be viewed as micro-wormhole solutions puncturing the BH horizon. Intriguingly, we may speculate that Kac-Moodions can retain interesting connections with the ER=EPR proposal, which relates wormholes solutions in the bulk with the entanglement of particles on the boundary \cite{Maldacena:2013xja}.














\vspace{1cm}

{\large \bf Acknowledgments} 
\vspace{4mm}

\noindent
We thank Juan Maldacena and Yong-Shi Wu for interesting discussions and remarks on these subjects.


\begin{thebibliography}{99}

\bibitem{Jackson:2014nla} 
  S.~Jackson, L.~McGough and H.~Verlinde,
  Nucl.\ Phys.\ B {\bf 901}, 382 (2015)
  doi:10.1016/j.nuclphysb.2015.10.013
  [arXiv:1412.5205 [hep-th]].

\bibitem{Turiaci:2017zwd}
  G.~Turiaci and H.~Verlinde,
  JHEP {\bf 1710} (2017) 167
  doi:10.1007/JHEP10(2017)167
  [arXiv:1701.00528 [hep-th]].

\bibitem{Mertens:2017mtv}
  T.~G.~Mertens, G.~J.~Turiaci and H.~L.~Verlinde,
  JHEP {\bf 1708} (2017) 136
  doi:10.1007/JHEP08(2017)136
  [arXiv:1705.08408 [hep-th]].


\bibitem{Strominger:2001pn}
  A.~Strominger,
  JHEP {\bf 0110} (2001) 034
  doi:10.1088/1126-6708/2001/10/034
  [hep-th/0106113].
  
    

\bibitem{Maldacena:1997re}
  J.~M.~Maldacena,
  Int.\ J.\ Theor.\ Phys.\  {\bf 38} (1999) 1113
   [Adv.\ Theor.\ Math.\ Phys.\  {\bf 2} (1998) 231]
  doi:10.1023/A:1026654312961, 10.4310/ATMP.1998.v2.n2.a1
  [hep-th/9711200].
  
 

\bibitem{Sen:2002nu}
  A.~Sen,
  JHEP {\bf 0204} (2002) 048
  doi:10.1088/1126-6708/2002/04/048
  [hep-th/0203211].
  
\bibitem{Sen:2002in}
  A.~Sen,
  JHEP {\bf 0207} (2002) 065
  doi:10.1088/1126-6708/2002/07/065
  [hep-th/0203265].


\bibitem{Gutperle:2002ai}
  M.~Gutperle and A.~Strominger,
  JHEP {\bf 0204} (2002) 018
  doi:10.1088/1126-6708/2002/04/018
  [hep-th/0202210].

\bibitem{Strominger:2002pc}
  A.~Strominger,
  Conf.\ Proc.\ C {\bf 0208124} (2002) 20
  [hep-th/0209090].
  
\bibitem{Hashimoto:2002sk}
  K.~Hashimoto, P.~M.~Ho and J.~E.~Wang,
  Phys.\ Rev.\ Lett.\  {\bf 90} (2003) 141601
  doi:10.1103/PhysRevLett.90.141601
  [hep-th/0211090].
  
\bibitem{Ohta:2003uw}
  N.~Ohta,
  Phys.\ Lett.\ B {\bf 558} (2003) 213
  doi:10.1016/S0370-2693(03)00274-0
  [hep-th/0301095].

\bibitem{Strominger:2001pn}
  A.~Strominger,
  JHEP {\bf 0110} (2001) 034
  doi:10.1088/1126-6708/2001/10/034
  [hep-th/0106113].
  
  \bibitem{Mandal:2017thl}
  G.~Mandal, P.~Nayak and S.~R.~Wadia,
  JHEP {\bf 1711} (2017) 046
  doi:10.1007/JHEP11(2017)046
  [arXiv:1702.04266 [hep-th]].


\bibitem{Ponsot:1999uf}
  B.~Ponsot and J.~Teschner,
  hep-th/9911110.
  
\bibitem{zoto} 
D.~Bagrets, A.~Altland and A.~Kamenev, 
Nucl. Phys. {\bf B} 911, 191 (2016) arXiv:1607.00694 [cond-mat.str-el]; 
arXiv:1702.08902 [cond-mat.str-el].  
  
\bibitem{Dorn:1994xn}
  H.~Dorn and H.~J.~Otto,
  Nucl.\ Phys.\ B {\bf 429} (1994) 375
  doi:10.1016/0550-3213(94)00352-1
  [hep-th/9403141].
  
\bibitem{Zamolodchikov:1995aa}
  A.~B.~Zamolodchikov and A.~B.~Zamolodchikov,
  Nucl.\ Phys.\ B {\bf 477} (1996) 577
  doi:10.1016/0550-3213(96)00351-3
  [hep-th/9506136].
  
\bibitem{Teschner:1995yf}
  J.~Teschner,
  Phys.\ Lett.\ B {\bf 363} (1995) 65
  doi:10.1016/0370-2693(95)01200-A
  [hep-th/9507109].
  
\bibitem{T1}
R.M.~Kashaev, 
Lett. Math. Phys. {\bf 43}, 105 (1998).

\bibitem{T2}
T.~Dimofte, 
Adv. Theor. Math. Phys. {\bf 17}, 479 (2013) [arXiv:1102.4847 [hep-th]].

\bibitem{T3}
J.~Teschner and G.~Vartanov,  
Lett. Math. Phys. {\bf 104}, 527 (2014) [arXiv:1202.4698 [hep-th]].

\bibitem{Lo:1993hp}
  H.~K.~Lo and J.~Preskill,
  Phys.\ Rev.\ D {\bf 48} (1993) 4821
  doi:10.1103/PhysRevD.48.4821
  [hep-th/9306006].

\bibitem{Wilczek:1989kn}
  F.~Wilczek and Y.~S.~Wu,
  Phys.\ Rev.\ Lett.\  {\bf 65} (1990) 13.
  

\bibitem{Bais:1998yn} 
  F.~A.~Bais and N.~M.~Muller,
  Nucl.\ Phys.\ B {\bf 530}, 349 (1998)
  doi:10.1016/S0550-3213(98)00572-0
  [hep-th/9804130].

\bibitem{Bais:2002ye} 
  F.~A.~Bais, N.~M.~Muller and B.~J.~Schroers,
  Nucl.\ Phys.\ B {\bf 640}, 3 (2002)
  doi:10.1016/S0550-3213(02)00572-2
  [hep-th/0205021].

\bibitem{Marciano:2010jm} 
  A.~Marciano,
  doi:10.1142/9789814374552\_0494
  arXiv:1003.0395 [hep-th].


\bibitem{Marciano:2010gq} 
  A.~Marciano, G.~Amelino-Camelia, N.~R.~Bruno, G.~Gubitosi, G.~Mandanici and A.~Melchiorri,
  JCAP {\bf 1006}, 030 (2010)
  doi:10.1088/1475-7516/2010/06/030
  [arXiv:1004.1110 [gr-qc]].

\bibitem{Ballesteros:2014kaa} 
  A.~Ballesteros, F.~J.~Herranz, C.~Meusburger and P.~Naranjo,
  SIGMA {\bf 10}, 052 (2014)
  doi:10.3842/SIGMA.2014.052
  [arXiv:1403.4773 [math-ph]].
  
\bibitem{S1}  
G.~'t Hooft, 
Commun. Math. Phys. {\bf 117} (1988) 685-700.

\bibitem{S2} 
S. Deser and R. Jackiw, Classical and quantum scattering on a cone, Commun. Math. Phys. 118 (1988) 495-509.  

\bibitem{S3}
Ph. de Sousa Gerbert and R. Jackiw, 
Commun. Math. Phys. {\bf 124} (1988) 229-260.
  
\bibitem{Shenker:2013pqa}
  S.~H.~Shenker and D.~Stanford,
  JHEP {\bf 1403} (2014) 067
  doi:10.1007/JHEP03(2014)067
  [arXiv:1306.0622 [hep-th]].
  
  
\bibitem{Banados:1998tb}
  M.~Banados, T.~Brotz and M.~E.~Ortiz,
  Phys.\ Rev.\ D {\bf 59} (1999) 046002
  doi:10.1103/PhysRevD.59.046002
  [hep-th/9807216].
  
\bibitem{12}
P. Goddard, A. Kent and D. Olive, Comm. Math. Phys. {\bf 103}, 105 (1986).
  
\bibitem{Addazi:2015gna}
  A.~Addazi,
  Int.\ J.\ Mod.\ Phys.\ A {\bf 32} (2017) no.16,  1750087
  doi:10.1142/S0217751X17500877
  [arXiv:1508.04054 [gr-qc]].
  
\bibitem{Addazi:2016cad}
  A.~Addazi,
  Int.\ J.\ Geom.\ Meth.\ Mod.\ Phys.\  {\bf 13} (2016) no.06,  1650082
  doi:10.1142/S0219887816500821
  [arXiv:1603.08719 [gr-qc]].
  
\bibitem{Addazi:2015hpa}
  A.~Addazi,
  Springer Proc.\ Phys.\  {\bf 208} (2018) 115
  [arXiv:1510.05876 [gr-qc]].
  
\bibitem{Addazi:2015cho}
  A.~Addazi,
  Electron.\ J.\ Theor.\ Phys.\  {\bf 12} (2015) no.34,  89.
  
\bibitem{Addazi:2018cyn}
  A.~Addazi and A.~Marciano,
  arXiv:1802.09940 [hep-th].
  
\bibitem{Addazi:2017qwt}
  A.~Addazi and A.~Marciano,
  Sci.\ China Phys.\ Mech.\ Astron.\  {\bf 61} (2018) no.12,  120421
  doi:10.1007/s11433-018-9263-0
  [arXiv:1707.05347 [hep-th]].
  
  
\bibitem{Addazi:2017xur}
  A.~Addazi, P.~Chen, A.~Marciano and Y.~S.~Wu,
  arXiv:1707.00347 [hep-th].
  
\bibitem{Maldacena:2013xja}
  J.~Maldacena and L.~Susskind,
  Fortsch.\ Phys.\  {\bf 61} (2013) 781
  doi:10.1002/prop.201300020
  [arXiv:1306.0533 [hep-th]].



\end{thebibliography}
\end{document}